\documentclass[sigplan,screen]{acmart}

\usepackage{textcomp}
\usepackage{hyperref}

\usepackage{todonotes}
\newcommand{\TODO}[1]{\todo[inline]{#1}}

\setcopyright{none}
\renewcommand\footnotetextcopyrightpermission[1]{}

\usepackage{longtable}
\usepackage{fancyvrb}

\definecolor{friendlybg}{HTML}{f0f0f0}
\definecolor{lightgray}{rgb}{.9,.9,.9}
\definecolor{darkgray}{rgb}{.4,.4,.4}
\definecolor{purple}{rgb}{0.65, 0.12, 0.82}

\acmConference[MLCommons Science Working Group Report]{MLCommons CloudMask Benchmark with Early Stopage }{Jun. 30,  revised 7 Dec. 2023}{Charlottesville, VA}

\begin{document}

\newcommand{\TITLE}{An Overview of MLCommons Cloud Mask Benchmark: \\ Related Research and Data \\ {\normalsize Version 1.0}}

\title[Overview of MLCommons Cloud Mask: Related Research]{\TITLE}
\titlenote{\url{https://github.com/laszewski/papers/raw/master/vonLaszewski-cloudmask-related.pdf}}

\author{Gregor von Laszewski}
\email{laszewski@gmail.com}
\orcid{0000-0001-9558-179X}
\authornote{MLCommons authorized submitting author}
\affiliation{%
  \institution{University of Virginia}
  \streetaddress{Biocomplexity Institute and Initiative\\
Town Center Four\\
994 Research Park Boulevard}
  \city{Charlottesville}
  \state{VA}
  \postcode{22911}
  \country{USA}
}

\author{Ruochen Gu}
\email{bill.ruochen.gu@gmail.com}
\affiliation{%
  \institution{}
  \streetaddress{}
  \city{Shanghai}
  \state{}
  \postcode{}
  \country{CN}
}

\renewcommand{\shortauthors}{von Laszewski et al.}

\begin{abstract}

  Cloud masking is a crucial task that is well-motivated for
  meteorology and its applications in environmental and atmospheric
  sciences. Its goal is, given satellite images, to accurately
  generate cloud masks that identify each pixel in image to contain
  either cloud or clear sky.  In this paper, we summarize some of the
  ongoing research activities in cloud masking, with a focus on the
  research and benchmark currently conducted in MLCommons Science
  Working Group.  This overview is produced with the hope that others
  will have an easier time getting started and collaborate on the
  activities related to MLCommons Cloud Mask Benchmark.

\end{abstract}

\ccsdesc[500]{Applied computing~Earth and atmospheric sciences}
\ccsdesc[500]{Computing methodologies~Image representations}
\ccsdesc[500]{Information systems~Test collections}
\ccsdesc[500]{Hardware~Testing with distributed and parallel systems}

\keywords{cloudmask, cloudmesh, datasets, MLCommons, benchmark}

\received[Version from]{8. June, revised 7 December, 2023}

\settopmatter{printfolios=true}
\maketitle

\section{MLCommons Cloud Mask Activities}

\nocite{las-2023-ai-workflow} 

The Cloud Mask Benchmark is part of the research currently conducted
by MLCommons \cite{Farrell2021MLPerfHA} Science Working Group
\cite{www-mlcommons-science-github}.  We hope that others will
contribute to this document to enhance its scope. Please contact
Gregor von Laszewski (laszewski@gmail.com) so that we can coordinate
with you.

As of this moment, we are aware of several activities regarding
MLCommons Cloud Mask Benchmark.

\begin{enumerate}
  
\item The original benchmark was contributed by Samuel Jackson and
  Juri Papaya \cite{Thiyagalingam2022AIBF,jackson-2020-eu} from
  Rutherford Labs.  The reference implementation is based on U-Net
  \cite{RFB15a}.

\item A cloud mask benchmark activity by Junji Yin on PEARL
  \cite{Thiyagalingam2022AIBF}.

\item A number of activities carried out by Gregor von Laszewski on
  Rivanna, University of Virginia's High Performance Computing
  Cluster. This activity contains significant contributions:

    \begin{enumerate} 

    \item Introduction of a README to showcase how to run the code
      that has been reused and modified successfully by others.
      
    \item Introduction of target directories that showcase how to use
      templates to run cloudmask benchmarks on various HPC machines,
      DGX station, and a Linux desktop with an NVIDIA card.

    \item Introduction of enhanced timers to measure execution time
      for different parts of the benchmark program.

    \item Usage of Cloudmesh StopWatch to provide easy human readable
      timers that can be parsed with little effort through exports as
      CSV data.

    \item Development and usage of a hyper-parameter permutation
      framework that enables the cloud mask model to be experimented
      with different hyper-parameters, including epochs, batch sizes,
      learning rates, etc. This work is also reused in other MLCommons
      Science Benchmarks \cite{las22-cloudmesh-cc-reu}. The work
      simplies benchmark results following the FAIR principal while
      integrating the hyperparameters as metadata.

    \item Development of a workflow system that enables the use of
      hybrid compute resources through templates
      \cite{las-2023-escience-cloudmask}.

    \item Application of the aforementioned work to education
      \cite{las-2023-mlcommons-edu-eq}.

    \item Hosting of a development repository for MLCommons Cloud Mask
      Benchmark code base, as part of MLCommons Science Working Group
      \cite{github-laszewsk-mlcommons}.

    \item Execution of a substantial number of benchmark experiments.

\end{enumerate}

\item A number of activities by New York University ("NYU") AI for
  Scientific Research (AIFSR) Benchmark Team on Greene, NYU High
  Performance Computing Cluster.

  \begin{enumerate}
    
  \item Modification of reference implementation to include early
    stoppage into model training, built onto the activities from
    Rutherford Labs and UVA.

  \item Implementation of a new accuracy metric introduced by Samuel
    Jackson, Juri Papaya, and Gregor von Laszewski.

  \item Coordinated the benchmark experiments with bash script that
    replicates a small number of features from the previous more
    comprehensive activity conducted by Gregor von Laszewski.

        \end{enumerate}

        NYU AIFSR's benchmark activity contains a limited number of
        experiments in contrast to the activity done by Gregor von
        Laszewski. A joint report of both efforts is under
        preparation. Several versions of the report were started such
        as \cite{las23-cloudmask}. The latest report is not yet
        available.

\end{enumerate}

\section{Overview of Cloud Mask and its Related Work}

Since last century, several methods have been developed for cloud
masking, ranging from rule-based techniques
\cite{Saunders1986AnAS,Saunders1988AnIM,Merchant2005ProbabilisticPB,
  Zhu2012ObjectbasedCA} to modern deep learning approaches
\cite{Li2019DeepLB,Domnich2021KappaMaskAC,Yan2018CloudAC,WIELAND2019111203,JEPPESEN2019247}. Among
the more traditional, rule-based techniques, two popular methodologies
have been threshold cloud screening
\cite{Saunders1986AnAS,Saunders1988AnIM} and Bayesian cloud masking
\cite{Merchant2005ProbabilisticPB}.

Threshold screening methods consist of several threshold tests where
spectral and spatial properties of satellite images are compared with
those ranges that are believed to indicate the presence of a clear sky
pixel. And those other pixels that are not labeled as clear sky are
then flagged as cloudy. This school of methodologies were widely used
from the late 1980s to the early 2000s
\cite{Merchant2005ProbabilisticPB}.

The gradual transition away from threshold screening methods was due
to its long-criticized limitations: firstly, threshold settings rely
heavily on domain expertise about indicators of cloudiness that may
not be objective, which also makes later modification and updates
difficult; secondly, thresholds provide users little flexibility in
the trade-off between coverage and accuracy; third, threshold tests do
not make use of all available prior information. These shortcomings of
threshold screening methods are improved by later developed Bayesian
methods \cite{Merchant2005ProbabilisticPB}.

The Bayesian approach applies Bayes' theorem on prior meteorology
information to deduce for each pixel the probability of containing
cloud or clear sky, and thereafter generating a cloud mask as
output. As a result, these Bayesian approaches are fully probabilistic
and make good use of prior information. Compared to threshold tests,
Bayesian methods achieve better accuracy in predicting pixels'
cloudiness, offering generality and conceptual clarity in its
approach, and enhancing maintainability and adaptability largely
\cite{Merchant2005ProbabilisticPB}.

More recently, the rising popularity of deep learning has led to the
use of CNNs for generating cloud masks. Deep learning methods
\cite{Li2019DeepLB,Domnich2021KappaMaskAC,Yan2018CloudAC,WIELAND2019111203,JEPPESEN2019247}
use computer vision models (CNNs) and treat the cloud mask task as
that of image segmentation tasks. CNNs have achieved superior
performance thanks to their automatic feature extraction ability. A
research paper published in 2019 \cite{JEPPESEN2019247} introduces
Remote Sensing Network (RS-Net), which is a CNN architecture branched
out of U-Net \cite{RFB15a} for cloud masking and was shown to achieve
higher performance compared to the state-of-the-art rule-based
approach known as Fmask \cite{Zhu2012ObjectbasedCA}. KappaMask
\cite{Domnich2021KappaMaskAC} and MSCFF \cite{Li2019DeepLB} are two
additional U-Net based CNN model that outperformed Fmask. All these
models have reported their performances on several satellite images
such as Sentinel-2, Landsat, etc., and also made use of
human-annotated (some assisted by software) ground truth values (See
in Table \ref{tab:datasets}). On the other hand, MLCommons Cloud Mask
Benchmark operates on SLSTR images from the newer Sentinel-3
satellite, which uses Bayesian approach generated cloud masks as
ground truth. The reference implementation provided by MLCommons
Science Working Group achieved 92\% classification accuracy on the
Sentinel-3 test set \cite{Thiyagalingam2022AIBF}.

The aforementioned deep learning approaches towards cloud masking are
by no means exhaustive. If you know about other significant cloud
masking or deep learning approaches, please inform us and we will add
them here.

\begin{table*}[htb]
    \centering
    \caption{This table presents the several methods used for cloud masking with their respective dataset, ground truth, and performance.}
    \label{tab:datasets}
    \resizebox{1.8\columnwidth}{!}{
    \begin{tabular}{|l|c|l|l|l|l|}
    \hline
        {\bf} & {\bf Reference} & {\bf Dataset} & {\bf Ground-truth} & {\bf Model}  & {\bf Accuracy} \\ \hline
        1 & \cite{Merchant2005ProbabilisticPB} & ATSR-2 & Human annotation & Bayesian screening & 0.917\\ \hline
        2 & \cite{WIELAND2019111203} & Sentinel-2 & Software-assisted human annotation (QGIS) & U-Net & 0.90 \\ \hline
        3 & \cite{WIELAND2019111203} & Landsat TM & Software-assisted human annotation (QGIS) & U-Net & 0.89 \\ \hline
        4 & \cite{WIELAND2019111203} & Landsat ETM+ & Software-assisted human annotation (QGIS) & U-Net & 0.89 \\ \hline
        5 & \cite{WIELAND2019111203} & Landsat OLI & Software-assisted human annotation (QGIS) & U-Net & 0.91 \\ \hline
        6 & \cite{Li2019DeepLB} & GaoFen-1 & Human annotation & MFFSNet & 0.98, mIOU = 0.87 \\ \hline
        7 & \cite{Domnich2021KappaMaskAC} & Sentinel 2 & Software-assisted human annotation (CVAT) & KappaMask & 0.91 \\ \hline
        8 & \cite{JEPPESEN2019247} & Landsat 8 Biome and SPARCS & Human annotation & RS-Net & 0.956 \\ \hline
    \end{tabular}}

\end{table*}

\section{Dataset} 

For MLCommons Cloud Mask Benchmark, we use the satellite images from Sentinel-3. 

\subsection{Sentinel-3}

According to \cite{Sentinel84:online}
``Sentinel-3 is an ocean and land mission composed of two identical satellites (Sentinel-3A and Sentinel-3B).''

Sentinel-3 makes use of multiple sensing instruments to accomplish its objectives:

\begin{itemize}
    \item SLSTR (Sea and Land Surface Temperature Radiometer)
\item OLCI (Ocean and Land Colour Instrument)
\item SRAL (SAR Altimeter)
\item DORIS (Doppler Orbitography and Radiopositioning Integrated by Satellite)
\item MWR (Microwave Radiometer).

\end{itemize}

``SLSTR and OLCI are optical instruments that are used to provide data continuity for ENVISAT's AATSR and MERIS instruments and the swaths of the two instruments overlap, allowing for new combined applications. OLCI is a medium-resolution imaging spectrometer, using five cameras to provide a wide field of view.
SRAL, DORIS, MWR and LRR are used for topographic measurements of the ocean and inland water.'' \cite{Sentinel84:online}

\begin{figure}[htb]
\centering\includegraphics[width=0.8\columnwidth]{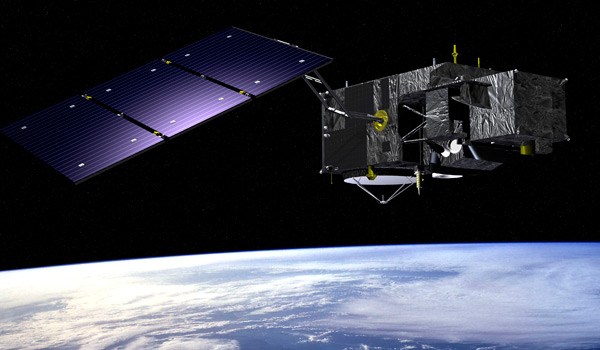}
\caption{A Sentinel-3 Satelite.}
\label{fig:sat}
\end{figure}

One of the satellites is shown in Figure~\ref{fig:sat}. The Mission
Orbit is sun-synchronous, set at a height of 814.5km with an
inclination of 98.65$^{o}$ and a repeat cycle of 27 days
\cite{Sentinel84:online}.

\subsection{MLCommons Cloud Mask Dataset}

MLCommons Cloud Mask Benchmark uses 180GB of satellite images from
Sentinel-3 SLSTR (Level-1 processing, TOA Radiances and Brightness
Temperature) satellite images. The dataset consists of 1070 images,
captured during days and nights. The dataset also includes a cloud
mask for each image, generated using Bayesian techniques. The
reference implementation uses these cloud masks as ground truths for
training and testing.

The dataset comes with the train-test split, where 970 images are used
for training, and 100 images are used for testing.  The images are of
the dimension $1200 \times 1500$ with 15 different channels and 1
channel of Bayesian mask. Among the 15 channels, 3 channels are used
to represent brightness, 6 channels are used to represent reflectance,
and the remaining 6 channels are used to represent radiance. However,
for the provided reference implementation, only a total of 10
channels, i.e., 6 channels of reflectance, 3 channels of brightness,
and 1 channel of Bayesian mask are used as model inputs for training
and testing.

\begin{figure*}[htb]

\centering\includegraphics[width=0.75\textwidth]{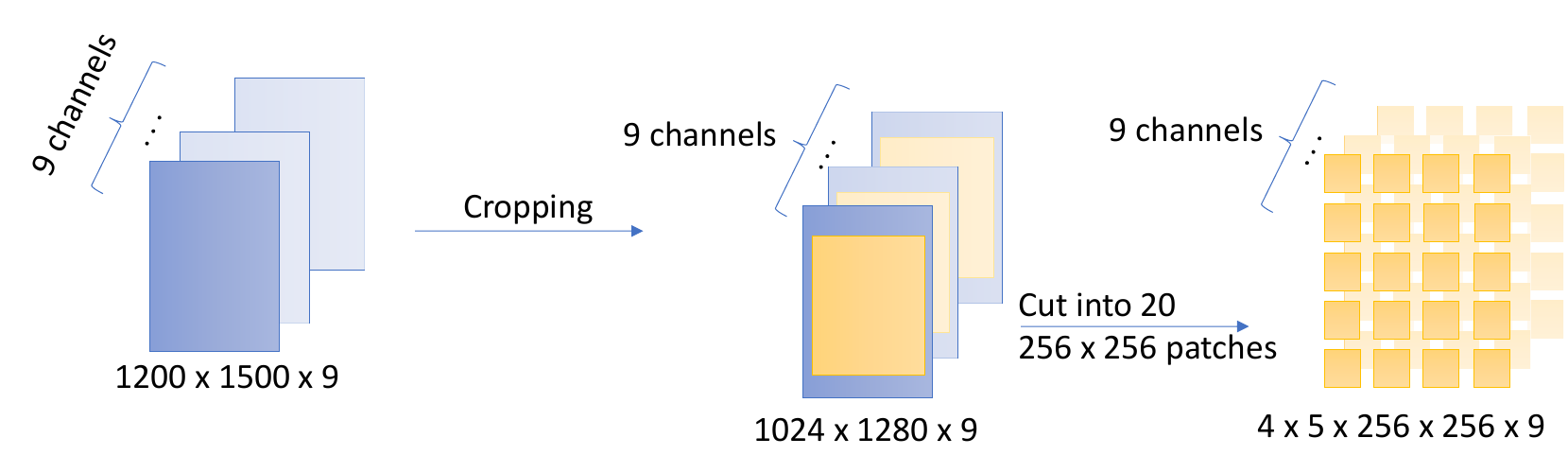}
\caption{The preprocessing of the training data.}
\label{fig:preprocessing-training}

\bigskip

\centering\includegraphics[width=0.75\textwidth]{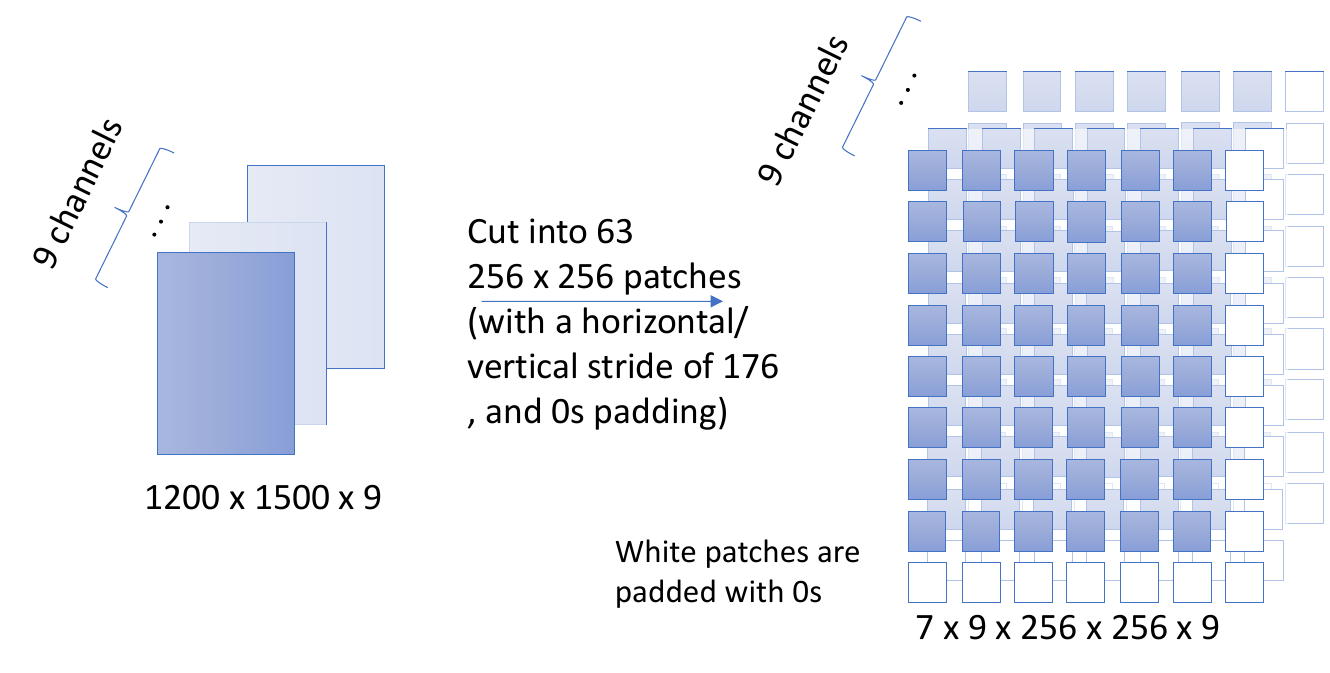}
\caption{The preprocessing of testing data}
\label{fig:preprocessing-testing}

\end{figure*}

\subsection{Data Loading and Preprocessing} \label{Preprocessing}

For training data preprocessing, the images are first cropped from the
dimension of $1200 \times 1500 \times 9$ to
$1024 \times 1280 \times 9$ and then divided into 20 smaller-sized
$256 \times 256 \times 9$ patches. After creating these patches out of
each image in training set, we get a total of $19400$ patches for
training. These patches are further split into training and validation
set with $80/20$ split ratio, and then sent for training after
shuffling.

For the test dataset, the images are neither cropped nor
shuffled. Instead, each test image are cut into 63 smaller patches of
dimension $256 \times 256 \times 9$, by applying a horizontal and
vertical stride of 176 pixels with zeros padding on the right and
bottom edges of each image. We then get a total of $6300$ patches for
entire test dataset. After getting the predictions from the model,
these $256 \times 256 \times 1$ output patches (predicted cloud mask)
are reconstructed to the size of $1200 \times 1500 \times 1$ and then
evaluated with the Bayesian mask ground truth that has the same
dimension. This preprocessing pipelines for training and testing are
shown in Figure \ref{fig:preprocessing-training} and Figure
\ref{fig:preprocessing-testing}.

\subsection{Training}

During training, the model takes a preprocessed patch of dimension
$256 \times 256 \times 9$, and generates a cloud mask of dimension
$256 \times 256 \times 1$. Once the cloud masks have been generated by
the model during training, the accuracy is reported as the percentage
of total pixels that are correctly classified compared to ground
truth.

\subsection{Testing}

During testing, the model generates a cloud mask of dimension
$256 \times 256 \times 1$ for each $256 \times 256 \times 9$
patch. For each pixel in the image, the model outputs the probability
of that pixel containing clear sky. Pixels that have a probability
higher than 50\% are labeled as clear sky, and cloudy otherwise. Then,
those patches are then reconstructed back to full-size masks of
dimension $1200 \times 1500 \times 1$.

\begin{figure*}[htb]
\centering\includegraphics[width=0.75\textwidth]{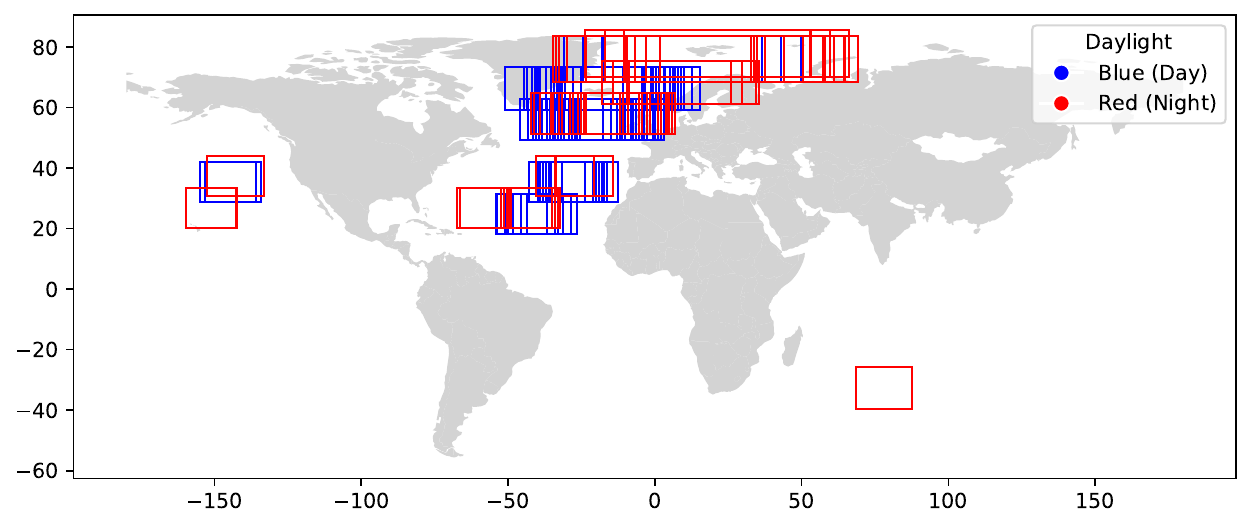}
\caption{The location of the satellite images represented as frames used for inference.}
\label{fig:frames-inference}

\centering\includegraphics[width=0.8\textwidth]{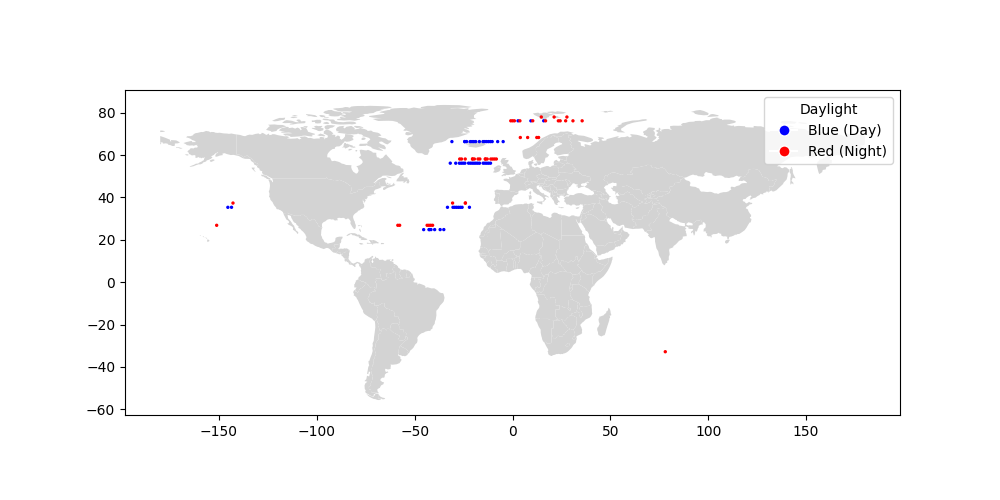}
\caption{The location of the center of the satellite images used for inference.}
\label{fig:frames-dot}
\end{figure*}

\begin{figure*}[htb]
\centering\includegraphics[width=0.8\textwidth]{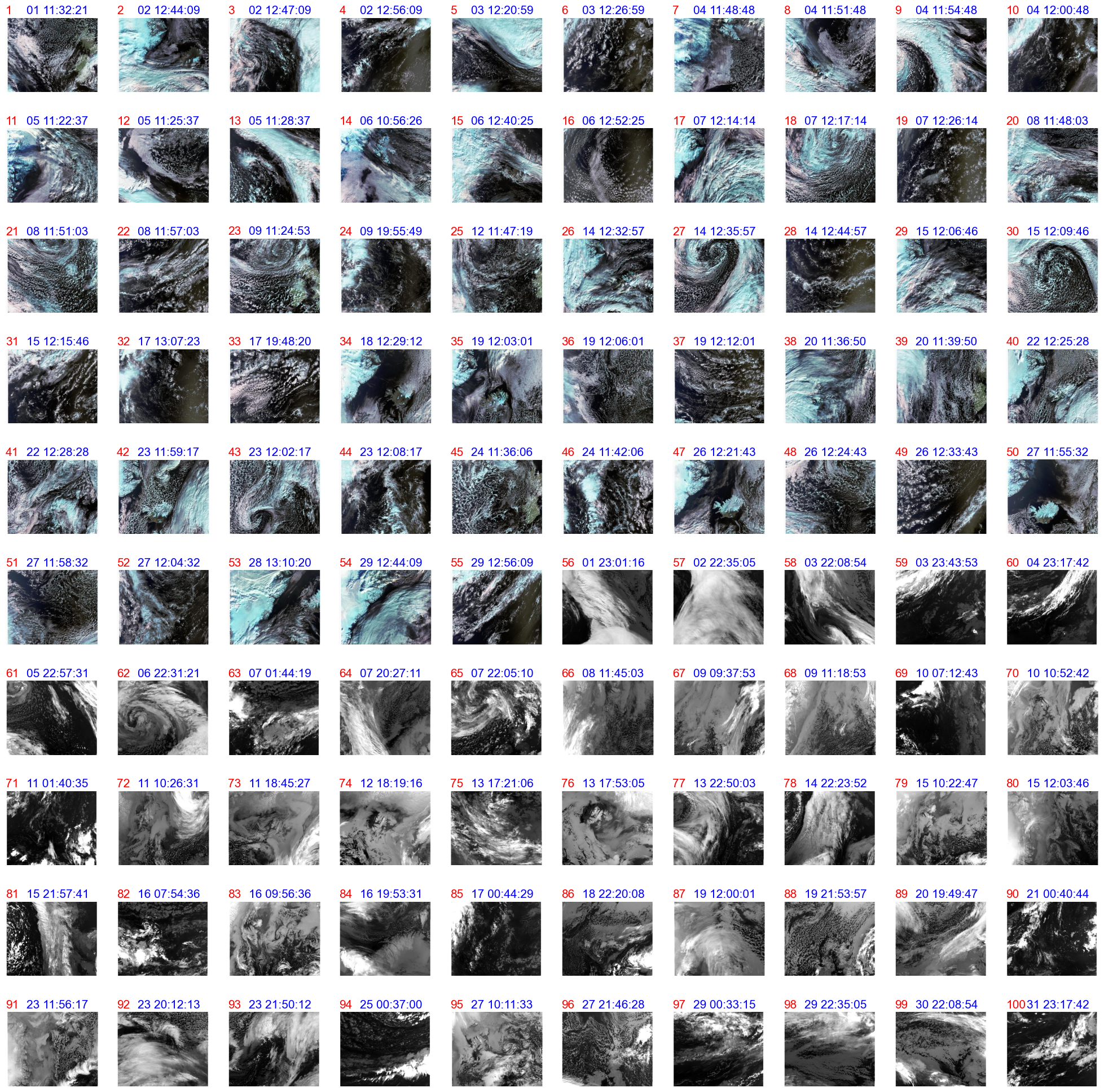}
\caption{The raw satellite images at the locations where inference is chosen.}
\label{fig:frames-raw}
\end{figure*}

\begin{figure*}[htb]
\centering\includegraphics[width=0.8\textwidth]{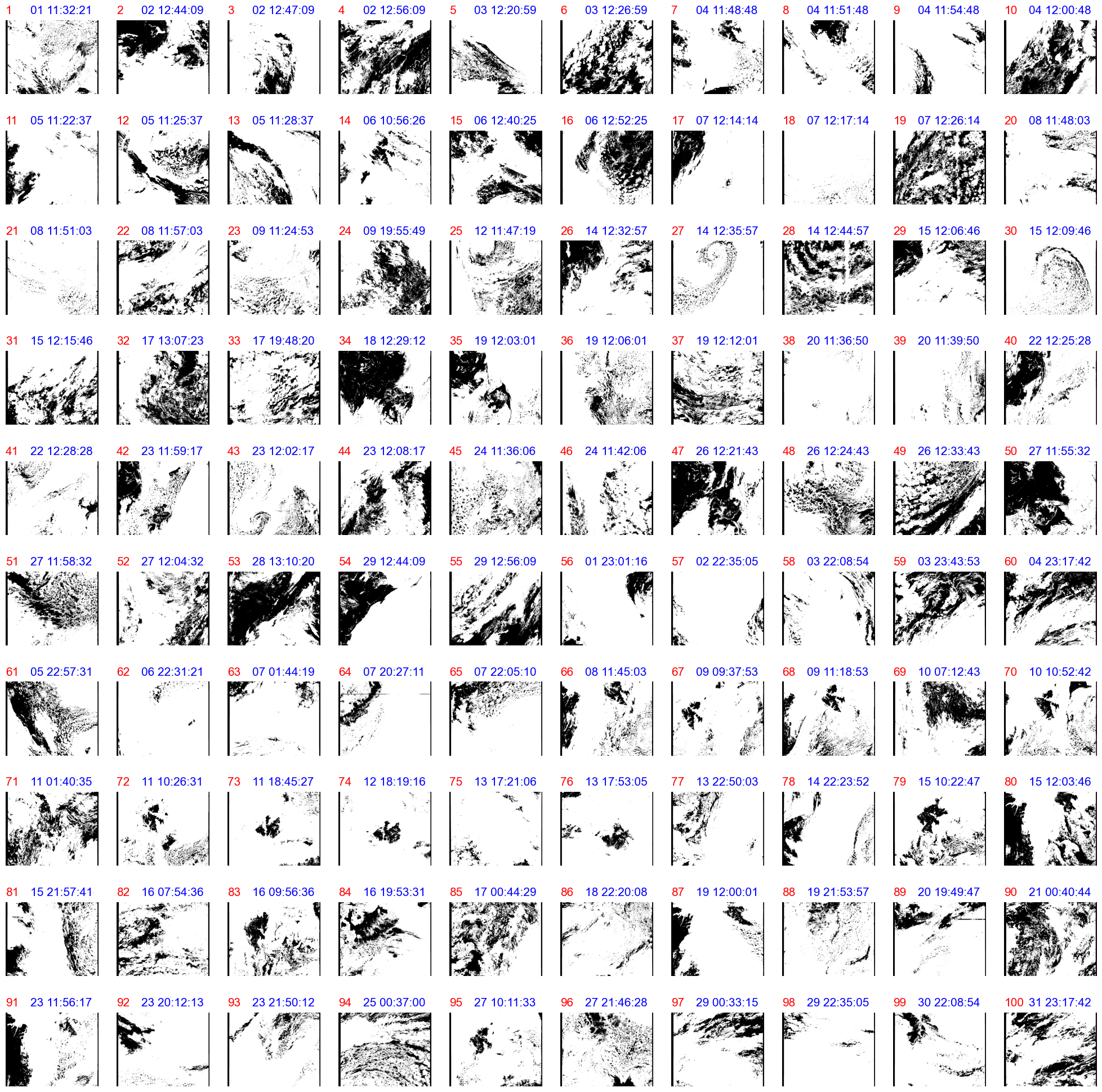}
\caption{The mask images at the locations where inference is chosen.}
\label{fig:frames-mask}
\end{figure*}

The locations of the images used in the testing are depicted in
Figure~\ref{fig:frames-inference} as well as their coordinate centers
in Figure~\ref{fig:frames-dot}. As one can observe from the figures,
most of the testing images are captured in the region of North
Altantic Ocean and of the West Coast of Europe.  Furthermore, we
display in Figure \ref{fig:frames-raw} the raw satellite images from
the Sentinel-3 database that reflect the locations where the testing
images are located. Figure \ref{fig:frames-mask} shows the cloud masks
of the testing images.  The Table \ref{tab:inference} shows the
individual attributes for the specific locations identified by a
counter.

\section{Conclusion}

In this paper, we provide a list of related activities under MLCommons
Cloud Mask Benchmark. The paper includes an overview of related
research in the field of cloud masking in general. In addition, the
paper provides a walk-through and illustration of the Sentinel-3
satellite image dataset used for MLCommons Cloud Mask Benchmark. With
this paper, we hope to enhance communication between activities on
this benchmark and support others to have an easier time getting
started with the MLCommons Cloud Mask benchmark.

\begin{acks}

  Work was in part funded by (a) NIST 60NANB21D151T (b) NSF
  CyberTraining: CIC: CyberTraining for Students and Technologies from
  Generation Z with the award numbers 1829704 and 2200409, and (c)
  Department of Energy under the grant Award No. DE-SC0023452 (d) NSF
  Collaborative Research: Framework: Software: CINES: A Scalable
  Cyberinfrastructure for Sustained Innovation in Network Engineering
  and Science with the award number 2210266. The work from the UVA
  team was conducted at the Biocomplexity Institute at
  the University of Virginia.  We like to thank the NYU AIFSR team for
  their contributions. We especially like to thank Ruochen Gu who
  continued to work on this project on voluntary basis.

\end{acks}


\bibliographystyle{ACM-Reference-Format}
\bibliography{vonLaszewski-cloudmask-related}


\section*{Contributions}

Ruochen Gu has conducted the work of identifying related research as a
student researcher from NYU AIFSR Benchmark Team. He continued this
work on a voluntary basis due to his interest in this project. {\em
  GvL} has contributed significantly to porting cloudmask onto
different machines while making the code portable and contributed the
cloudmesh-ee workflow code, the cloudmesh StopWatch, and integrated
the cloudmesh timers and logging, into the code. He ran all benchmarks
on Rivanna and the Desktop.  In discussions with Rutherford Lab, a new
accuracy value was introduced that was not included in the original
version distributed by MLCommons. He also facilitated many hackathons
with the NYU team. The work described here is cited in the NYU report.

\clearpage

\appendix

\onecolumn

\section{Table of Locations Used for Inference}

\begin{longtable}{rlllll}
\caption{Location instance ids used for inference}\label{tab:inference}\\
\toprule
 counter & daylight &          start\_time &           stop\_time &       creation\_date &       instance\_id \\
\midrule
\endfirsthead

\toprule
 Counter & Daylight &          Start Time &           Stop Time &       Creation Date &       Instance ID \\
\midrule
\endhead
\midrule
\multicolumn{6}{r}{{Continued on next page}} \\
\midrule
\endfoot

\bottomrule
\endlastfoot
       1 &      day & 2019-10-01 11:32:21 & 2019-10-01 11:35:21 & 2019-10-02 15:32:11 & 0179\_050\_023\_1980 \\
       2 &      day & 2019-10-02 12:44:09 & 2019-10-02 12:47:09 & 2019-10-03 18:06:38 & 0180\_050\_038\_1800 \\
       3 &      day & 2019-10-02 12:47:09 & 2019-10-02 12:50:09 & 2019-10-03 18:08:06 & 0179\_050\_038\_1980 \\
       4 &      day & 2019-10-02 12:56:09 & 2019-10-02 12:59:09 & 2019-10-03 18:12:26 & 0179\_050\_038\_2520 \\
       5 &      day & 2019-10-03 12:20:59 & 2019-10-03 12:23:59 & 2019-10-04 17:35:32 & 0179\_050\_052\_1980 \\
       6 &      day & 2019-10-03 12:26:59 & 2019-10-03 12:29:59 & 2019-10-04 17:38:22 & 0179\_050\_052\_2340 \\
       7 &      day & 2019-10-04 11:48:48 & 2019-10-04 11:51:48 & 2019-10-05 17:29:18 & 0179\_050\_066\_1620 \\
       8 &      day & 2019-10-04 11:51:48 & 2019-10-04 11:54:48 & 2019-10-05 17:30:23 & 0179\_050\_066\_1800 \\
       9 &      day & 2019-10-04 11:54:48 & 2019-10-04 11:57:48 & 2019-10-05 17:31:29 & 0180\_050\_066\_1980 \\
      10 &      day & 2019-10-04 12:00:48 & 2019-10-04 12:03:48 & 2019-10-05 17:33:44 & 0179\_050\_066\_2340 \\
      11 &      day & 2019-10-05 11:22:37 & 2019-10-05 11:25:37 & 2019-10-06 16:15:03 & 0179\_050\_080\_1620 \\
      12 &      day & 2019-10-05 11:25:37 & 2019-10-05 11:28:37 & 2019-10-06 16:16:11 & 0179\_050\_080\_1800 \\
      13 &      day & 2019-10-05 11:28:37 & 2019-10-05 11:31:37 & 2019-10-06 16:17:18 & 0179\_050\_080\_1980 \\
      14 &      day & 2019-10-06 10:56:26 & 2019-10-06 10:59:26 & 2019-10-07 15:33:22 & 0179\_050\_094\_1620 \\
      15 &      day & 2019-10-06 12:40:25 & 2019-10-06 12:43:25 & 2019-10-07 17:07:14 & 0179\_050\_095\_1800 \\
      16 &      day & 2019-10-06 12:52:25 & 2019-10-06 12:55:25 & 2019-10-07 17:11:48 & 0179\_050\_095\_2520 \\
      17 &      day & 2019-10-07 12:14:14 & 2019-10-07 12:17:14 & 2019-10-09 12:34:33 & 0179\_050\_109\_1800 \\
      18 &      day & 2019-10-07 12:17:14 & 2019-10-07 12:20:14 & 2019-10-09 12:35:40 & 0179\_050\_109\_1980 \\
      19 &      day & 2019-10-07 12:26:14 & 2019-10-07 12:29:14 & 2019-10-09 12:39:03 & 0179\_050\_109\_2520 \\
      20 &      day & 2019-10-08 11:48:03 & 2019-10-08 11:51:03 & 2019-10-09 16:23:06 & 0179\_050\_123\_1800 \\
      21 &      day & 2019-10-08 11:51:03 & 2019-10-08 11:54:03 & 2019-10-09 16:24:31 & 0179\_050\_123\_1980 \\
      22 &      day & 2019-10-08 11:57:03 & 2019-10-08 12:00:03 & 2019-10-09 16:27:35 & 0179\_050\_123\_2340 \\
      23 &      day & 2019-10-09 11:24:53 & 2019-10-09 11:27:53 & 2019-10-10 17:18:44 & 0180\_050\_137\_1980 \\
      24 &      day & 2019-10-09 19:55:49 & 2019-10-09 19:58:49 & 2019-10-11 01:18:36 & 0179\_050\_142\_2340 \\
      25 &      day & 2019-10-12 11:47:19 & 2019-10-12 11:50:19 & 2019-10-13 16:24:41 & 0179\_050\_180\_1980 \\
      26 &      day & 2019-10-14 12:32:57 & 2019-10-14 12:35:57 & 2019-10-15 17:27:25 & 0179\_050\_209\_1800 \\
      27 &      day & 2019-10-14 12:35:57 & 2019-10-14 12:38:57 & 2019-10-15 17:28:51 & 0179\_050\_209\_1980 \\
      28 &      day & 2019-10-14 12:44:57 & 2019-10-14 12:47:57 & 2019-10-15 17:32:52 & 0179\_050\_209\_2520 \\
      29 &      day & 2019-10-15 12:06:46 & 2019-10-15 12:09:46 & 2019-10-16 16:05:34 & 0179\_050\_223\_1800 \\
      30 &      day & 2019-10-15 12:09:46 & 2019-10-15 12:12:46 & 2019-10-16 16:07:00 & 0179\_050\_223\_1980 \\
      31 &      day & 2019-10-15 12:15:46 & 2019-10-15 12:18:46 & 2019-10-16 16:09:34 & 0179\_050\_223\_2340 \\
      32 &      day & 2019-10-17 13:07:23 & 2019-10-17 13:10:23 & 2019-10-18 18:14:27 & 0179\_050\_252\_2520 \\
      33 &      day & 2019-10-17 19:48:20 & 2019-10-17 19:51:20 & 2019-10-19 01:41:23 & 0180\_050\_256\_2340 \\
      34 &      day & 2019-10-18 12:29:12 & 2019-10-18 12:32:12 & 2019-10-19 17:26:27 & 0180\_050\_266\_1800 \\
      35 &      day & 2019-10-19 12:03:01 & 2019-10-19 12:06:01 & 2019-10-20 17:25:39 & 0179\_050\_280\_1800 \\
      36 &      day & 2019-10-19 12:06:01 & 2019-10-19 12:09:01 & 2019-10-20 17:26:59 & 0179\_050\_280\_1980 \\
      37 &      day & 2019-10-19 12:12:01 & 2019-10-19 12:15:01 & 2019-10-20 17:29:47 & 0179\_050\_280\_2340 \\
      38 &      day & 2019-10-20 11:36:50 & 2019-10-20 11:39:50 & 2019-10-21 15:59:04 & 0179\_050\_294\_1800 \\
      39 &      day & 2019-10-20 11:39:50 & 2019-10-20 11:42:50 & 2019-10-21 16:00:28 & 0179\_050\_294\_1980 \\
      40 &      day & 2019-10-22 12:25:28 & 2019-10-22 12:28:28 & 2019-10-23 16:23:16 & 0179\_050\_323\_1800 \\
      41 &      day & 2019-10-22 12:28:28 & 2019-10-22 12:31:28 & 2019-10-23 16:24:24 & 0179\_050\_323\_1980 \\
      42 &      day & 2019-10-23 11:59:17 & 2019-10-23 12:02:17 & 2019-10-24 18:15:37 & 0179\_050\_337\_1800 \\
      43 &      day & 2019-10-23 12:02:17 & 2019-10-23 12:05:17 & 2019-10-24 18:17:00 & 0179\_050\_337\_1980 \\
      44 &      day & 2019-10-23 12:08:17 & 2019-10-23 12:11:17 & 2019-10-24 18:19:32 & 0179\_050\_337\_2340 \\
      45 &      day & 2019-10-24 11:36:06 & 2019-10-24 11:39:06 & 2019-10-25 16:46:19 & 0180\_050\_351\_1980 \\
      46 &      day & 2019-10-24 11:42:06 & 2019-10-24 11:45:06 & 2019-10-25 16:48:55 & 0179\_050\_351\_2340 \\
      47 &      day & 2019-10-26 12:21:43 & 2019-10-26 12:24:43 & 2019-10-27 16:59:12 & 0179\_050\_380\_1800 \\
      48 &      day & 2019-10-26 12:24:43 & 2019-10-26 12:27:43 & 2019-10-27 17:00:26 & 0179\_050\_380\_1980 \\
      49 &      day & 2019-10-26 12:33:43 & 2019-10-26 12:36:43 & 2019-10-27 17:04:26 & 0180\_050\_380\_2520 \\
      50 &      day & 2019-10-27 11:55:32 & 2019-10-27 11:58:32 & 2019-10-28 17:10:44 & 0179\_051\_009\_1800 \\
      51 &      day & 2019-10-27 11:58:32 & 2019-10-27 12:01:32 & 2019-10-28 17:13:22 & 0179\_051\_009\_1980 \\
      52 &      day & 2019-10-27 12:04:32 & 2019-10-27 12:07:32 & 2019-10-28 17:15:46 & 0179\_051\_009\_2340 \\
      53 &      day & 2019-10-28 13:10:20 & 2019-10-28 13:13:20 & 2019-10-29 18:11:38 & 0179\_051\_024\_1800 \\
      54 &      day & 2019-10-29 12:44:09 & 2019-10-29 12:47:09 & 2019-10-30 17:20:48 & 0179\_051\_038\_1800 \\
      55 &      day & 2019-10-29 12:56:09 & 2019-10-29 12:59:09 & 2019-10-30 17:26:09 & 0179\_051\_038\_2520 \\
      56 &    night & 2019-10-01 23:01:16 & 2019-10-01 23:04:16 & 2019-10-03 02:22:22 & 0179\_050\_030\_0900 \\
      57 &    night & 2019-10-02 22:35:05 & 2019-10-02 22:38:05 & 2019-10-04 03:19:30 & 0179\_050\_044\_0900 \\
      58 &    night & 2019-10-03 22:08:54 & 2019-10-03 22:11:54 & 2019-10-05 03:04:45 & 0179\_050\_058\_0900 \\
      59 &    night & 2019-10-03 23:43:53 & 2019-10-03 23:46:53 & 2019-10-05 04:17:23 & 0179\_050\_059\_0540 \\
      60 &    night & 2019-10-04 23:17:42 & 2019-10-04 23:20:42 & 2019-10-06 04:17:26 & 0179\_050\_073\_0540 \\
      61 &    night & 2019-10-05 22:57:31 & 2019-10-05 23:00:31 & 2019-10-07 02:49:22 & 0179\_050\_087\_0900 \\
      62 &    night & 2019-10-06 22:31:21 & 2019-10-06 22:34:21 & 2019-10-08 02:13:01 & 0179\_050\_101\_0900 \\
      63 &    night & 2019-10-07 01:44:19 & 2019-10-07 01:47:19 & 2019-10-08 06:11:44 & 0179\_050\_103\_0360 \\
      64 &    night & 2019-10-07 20:27:11 & 2019-10-07 20:30:11 & 2019-10-09 15:52:46 & 0179\_050\_114\_1080 \\
      65 &    night & 2019-10-07 22:05:10 & 2019-10-07 22:08:10 & 2019-10-10 07:30:03 & 0179\_050\_115\_0900 \\
      66 &    night & 2019-10-08 11:45:03 & 2019-10-08 11:48:03 & 2019-10-09 16:21:44 & 0179\_050\_123\_1620 \\
      67 &    night & 2019-10-09 09:37:53 & 2019-10-09 09:40:53 & 2019-10-10 15:10:07 & 0179\_050\_136\_1620 \\
      68 &    night & 2019-10-09 11:18:53 & 2019-10-09 11:21:53 & 2019-10-10 17:17:17 & 0179\_050\_137\_1620 \\
      69 &    night & 2019-10-10 07:12:43 & 2019-10-10 07:15:43 & 2019-10-11 11:46:43 & 0179\_050\_149\_0540 \\
      70 &    night & 2019-10-10 10:52:42 & 2019-10-10 10:55:42 & 2019-10-11 16:05:48 & 0179\_050\_151\_1620 \\
      71 &    night & 2019-10-11 01:40:35 & 2019-10-11 01:43:35 & 2019-10-12 06:07:33 & 0179\_050\_160\_0360 \\
      72 &    night & 2019-10-11 10:26:31 & 2019-10-11 10:29:31 & 2019-10-12 15:25:04 & 0180\_050\_165\_1620 \\
      73 &    night & 2019-10-11 18:45:27 & 2019-10-11 18:48:27 & 2019-10-13 00:25:31 & 0179\_050\_170\_1260 \\
      74 &    night & 2019-10-12 18:19:16 & 2019-10-12 18:22:16 & 2019-10-13 23:09:19 & 0179\_050\_184\_1260 \\
      75 &    night & 2019-10-13 17:21:06 & 2019-10-13 17:24:06 & 2019-10-14 21:57:23 & 0179\_050\_197\_5400 \\
      76 &    night & 2019-10-13 17:53:05 & 2019-10-13 17:56:05 & 2019-10-14 22:28:12 & 0180\_050\_198\_1260 \\
      77 &    night & 2019-10-13 22:50:03 & 2019-10-13 22:53:03 & 2019-10-15 03:02:39 & 0179\_050\_201\_0900 \\
      78 &    night & 2019-10-14 22:23:52 & 2019-10-14 22:26:52 & 2019-10-16 01:57:10 & 0179\_050\_215\_0900 \\
      79 &    night & 2019-10-15 10:22:47 & 2019-10-15 10:25:47 & 2019-10-16 14:25:44 & 0179\_050\_222\_1620 \\
      80 &    night & 2019-10-15 12:03:46 & 2019-10-15 12:06:46 & 2019-10-16 16:05:31 & 0179\_050\_223\_1620 \\
      81 &    night & 2019-10-15 21:57:41 & 2019-10-15 22:00:41 & 2019-10-17 01:03:23 & 0179\_050\_229\_0900 \\
      82 &    night & 2019-10-16 07:54:36 & 2019-10-16 07:57:36 & 2019-10-17 12:39:59 & 0179\_050\_235\_0360 \\
      83 &    night & 2019-10-16 09:56:36 & 2019-10-16 09:59:36 & 2019-10-17 15:26:06 & 0179\_050\_236\_1620 \\
      84 &    night & 2019-10-16 19:53:31 & 2019-10-16 19:56:31 & 2019-10-18 01:29:29 & 0179\_050\_242\_1080 \\
      85 &    night & 2019-10-17 00:44:29 & 2019-10-17 00:47:29 & 2019-10-18 06:04:49 & 0179\_050\_245\_0360 \\
      86 &    night & 2019-10-18 22:20:08 & 2019-10-18 22:23:08 & 2019-10-20 03:22:33 & 0179\_050\_272\_0900 \\
      87 &    night & 2019-10-19 12:00:01 & 2019-10-19 12:03:01 & 2019-10-20 17:25:12 & 0179\_050\_280\_1620 \\
      88 &    night & 2019-10-19 21:53:57 & 2019-10-19 21:56:57 & 2019-10-21 01:47:57 & 0179\_050\_286\_0900 \\
      89 &    night & 2019-10-20 19:49:47 & 2019-10-20 19:52:47 & 2019-10-22 00:31:58 & 0179\_050\_299\_1080 \\
      90 &    night & 2019-10-21 00:40:44 & 2019-10-21 00:43:44 & 2019-10-22 05:13:08 & 0180\_050\_302\_0360 \\
      91 &    night & 2019-10-23 11:56:17 & 2019-10-23 11:59:17 & 2019-10-24 18:15:37 & 0180\_050\_337\_1620 \\
      92 &    night & 2019-10-23 20:12:13 & 2019-10-23 20:15:13 & 2019-10-25 01:48:37 & 0179\_050\_342\_1080 \\
      93 &    night & 2019-10-23 21:50:12 & 2019-10-23 21:53:12 & 2019-10-25 02:31:07 & 0179\_050\_343\_0900 \\
      94 &    night & 2019-10-25 00:37:00 & 2019-10-25 00:40:00 & 2019-10-26 05:49:24 & 0180\_050\_359\_0360 \\
      95 &    night & 2019-10-27 10:11:33 & 2019-10-27 10:14:33 & 2019-10-28 14:44:45 & 0179\_051\_008\_1620 \\
      96 &    night & 2019-10-27 21:46:28 & 2019-10-27 21:49:28 & 2019-10-29 01:32:17 & 0179\_051\_015\_0900 \\
      97 &    night & 2019-10-29 00:33:15 & 2019-10-29 00:36:15 & 2019-10-30 05:00:58 & 0179\_051\_031\_0360 \\
      98 &    night & 2019-10-29 22:35:05 & 2019-10-29 22:38:05 & 2019-10-31 01:57:42 & 0179\_051\_044\_0900 \\
      99 &    night & 2019-10-30 22:08:54 & 2019-10-30 22:11:54 & 2019-11-01 02:57:13 & 0179\_051\_058\_0900 \\
     100 &    night & 2019-10-31 23:17:42 & 2019-10-31 23:20:42 & 2019-11-02 03:46:13 & 0179\_051\_073\_0540 \\
\end{longtable}


\section{Naming Convention}

\url{https://sentinels.copernicus.eu/web/sentinel/user-guides/sentinel-3-slstr/naming-convention}

The file naming convention of SLSTR products
(\href{https://earth.esa.int/documents/247904/1964331/Sentinel-3_PDGS_File_Naming_Convention}{see
Sentinel-3 PDGS File Naming Convention for more details}) is identified
by the sequence of fields described below:

\emph{\textbf{MMM\_SL\_L\_TTTTTT\_yyyymmddThhmmss\_YYYYMMDDTHHMMSS\_YYYYMMDDTHHMMSS\_{[}instance
ID{]}\_GGG\_{[}class ID{]}.SEN3}}

where:

\begin{itemize}
\item
  \textbf{MMM}~is the mission ID:

  \begin{itemize}
  \item
    S3A = SENTINEL-3A
  \item
    S3B = SENTINEL-3B
  \item
    S3\_ = for both SENTINEL-3A and 3B
  \end{itemize}
\item
  \textbf{SL}~is the data source/consumer (SL = SLSTR)
\item
  \textbf{L}~is the processing level
(
  \begin{itemize}
  \item
    "0" for Level-0
  \item
    "1" for Level-1
  \item
    "2" for Level-2
  \item
    underscore "\_" if processing level is not applicable
  \end{itemize}
\item
  \textbf{TTTTTT}~is the data Type ID

  \begin{itemize}
  \item
    Level 0 SLSTR data:

    \begin{itemize}
    \item
      "SLT\_\_\_" = ISPs.
    \end{itemize}
  \item
    Level-1 SLSTR data:

    \begin{itemize}
    \item
      "RBT\_\_\_" = TOA Radiances and Brightness Temperature
    \item
      "RBT\_BW" = browse product derived from "RBT\_\_\_".
    \end{itemize}
  \item
    Level-2 SLSTR data:

    \begin{itemize}
    \item
      "WCT\_\_\_" = 2 and 3 channels SST for nadir and along track view
    \item
      "WST\_\_\_" = L2P sea surface temperature
    \item
      "LST\_\_\_" = land surface temp
    \item
      "FRP\_\_\_" = Fire Radiative Power
    \item
      "WST\_BW" = browse product derived from "WST\_\_\_"
    \item
      "LST\_BW" = browse product derived from "LST\_\_\_".
    \end{itemize}
  \end{itemize}
\item
  \textbf{yyyymmddThhmmss}~is the sensing start time
\item
  \textbf{YYYYMMDDTHHMMSS}~is the sensing stop time
\item
  \textbf{YYYYMMDDTHHMMSS}~is the product creation date
\item
  \textbf{{[}instance ID{]}~}consists of 17 characters, either uppercase
  letters or digits or underscores "\_".
\end{itemize}

\begin{quote}
The instance id fields include the following cases, applicable as
indicated:

\begin{enumerate}
\item Instance ID for the instrument data products disseminated in
"stripes":

Duration,"\_", cycle number, "\_", relative orbit number,"\_", 4
underscores "\_", i.e.\\
DDDD\_CCC\_LLL\_\_\_\_\_\\
\item Instance ID for the instrument data products ~\\
disseminated in "frames":\\
Duration, "\_", cycle number, "\_", relative orbit number, "\_", frame
along track coordinate, i.e.\\
DDDD\_CCC\_LLL\_FFFF\\
\item Instance ID for the instrument data products disseminated in
"tiles".\\
Two sub-cases are applicable:

a) tile covering the whole globe:\\
\hspace*{0.333em}\hspace*{0.333em}\hspace*{0.333em}\hspace*{0.333em}\hspace*{0.333em}
"GLOBAL\_\_\_\_\_\_\_\_\_\_\_"\\

b) tile cut according to specific\\
geographical criteria:\\
Tile Identifier\\
ttttttttttttttttt\\
\item Instance ID for auxiliary data:\\
17 underscores "\_"

\end{enumerate}
\end{quote}

\begin{itemize}
\item
  \textbf{GGG}~identifies the centre which generated the file
\item
  \textbf{{[}class ID{]}}~identifies the class ID for instrument data
  products with conventional sequence~\textbf{P\_XX\_NNN}~where:

  \begin{itemize}
  \item
    P indicates the platform (O for operational, F for reference, D for
    development, R for reprocessing)
  \item
    XX indicates the timeliness of the processing workflow (NR for NRT,
    ST for STC, NT for NTC)
  \item
    NNN indicates the baseline collection or data usage.
  \end{itemize}
\item
  \textbf{.SEN3}~is the filename extension
\end{itemize}

Example of filename:

\verb|S3A_SL_2_LST____20151229T095534_20151229T114422_20160102T150019_6528_064_365______LN2_D_NT_001.SEN3|

\section{Inference Files}

\begin{verbatim}
S3A_SL_1_RBT____20191001T113221_20191001T113521_20191002T153211_0179_050_023_1980_LN2_O_NT_003.hdf
S3A_SL_1_RBT____20191002T124409_20191002T124709_20191003T180638_0180_050_038_1800_LN2_O_NT_003.hdf
S3A_SL_1_RBT____20191002T124709_20191002T125009_20191003T180806_0179_050_038_1980_LN2_O_NT_003.hdf
S3A_SL_1_RBT____20191002T125609_20191002T125909_20191003T181226_0179_050_038_2520_LN2_O_NT_003.hdf
S3A_SL_1_RBT____20191003T122059_20191003T122359_20191004T173532_0179_050_052_1980_LN2_O_NT_003.hdf
S3A_SL_1_RBT____20191003T122659_20191003T122959_20191004T173822_0179_050_052_2340_LN2_O_NT_003.hdf
S3A_SL_1_RBT____20191004T114848_20191004T115148_20191005T172918_0179_050_066_1620_LN2_O_NT_003.hdf
S3A_SL_1_RBT____20191004T115148_20191004T115448_20191005T173023_0179_050_066_1800_LN2_O_NT_003.hdf
S3A_SL_1_RBT____20191004T115448_20191004T115748_20191005T173129_0180_050_066_1980_LN2_O_NT_003.hdf
S3A_SL_1_RBT____20191004T120048_20191004T120348_20191005T173344_0179_050_066_2340_LN2_O_NT_003.hdf
S3A_SL_1_RBT____20191005T112237_20191005T112537_20191006T161503_0179_050_080_1620_LN2_O_NT_003.hdf
S3A_SL_1_RBT____20191005T112537_20191005T112837_20191006T161611_0179_050_080_1800_LN2_O_NT_003.hdf
S3A_SL_1_RBT____20191005T112837_20191005T113137_20191006T161718_0179_050_080_1980_LN2_O_NT_003.hdf
S3A_SL_1_RBT____20191006T105626_20191006T105926_20191007T153322_0179_050_094_1620_LN2_O_NT_003.hdf
S3A_SL_1_RBT____20191006T124025_20191006T124325_20191007T170714_0179_050_095_1800_LN2_O_NT_003.hdf
S3A_SL_1_RBT____20191006T125225_20191006T125525_20191007T171148_0179_050_095_2520_LN2_O_NT_003.hdf
S3A_SL_1_RBT____20191007T121414_20191007T121714_20191009T123433_0179_050_109_1800_LN2_O_NT_003.hdf
S3A_SL_1_RBT____20191007T121714_20191007T122014_20191009T123540_0179_050_109_1980_LN2_O_NT_003.hdf
S3A_SL_1_RBT____20191007T122614_20191007T122914_20191009T123903_0179_050_109_2520_LN2_O_NT_003.hdf
S3A_SL_1_RBT____20191008T114803_20191008T115103_20191009T162306_0179_050_123_1800_LN2_O_NT_003.hdf
S3A_SL_1_RBT____20191008T115103_20191008T115403_20191009T162431_0179_050_123_1980_LN2_O_NT_003.hdf
S3A_SL_1_RBT____20191008T115703_20191008T120003_20191009T162735_0179_050_123_2340_LN2_O_NT_003.hdf
S3A_SL_1_RBT____20191009T112453_20191009T112753_20191010T171844_0180_050_137_1980_LN2_O_NT_003.hdf
S3A_SL_1_RBT____20191009T195549_20191009T195849_20191011T011836_0179_050_142_2340_LN2_O_NT_003.hdf
S3A_SL_1_RBT____20191012T114719_20191012T115019_20191013T162441_0179_050_180_1980_LN2_O_NT_003.hdf
S3A_SL_1_RBT____20191014T123257_20191014T123557_20191015T172725_0179_050_209_1800_LN2_O_NT_003.hdf
S3A_SL_1_RBT____20191014T123557_20191014T123857_20191015T172851_0179_050_209_1980_LN2_O_NT_003.hdf
S3A_SL_1_RBT____20191014T124457_20191014T124757_20191015T173252_0179_050_209_2520_LN2_O_NT_003.hdf
S3A_SL_1_RBT____20191015T120646_20191015T120946_20191016T160534_0179_050_223_1800_LN2_O_NT_003.hdf
S3A_SL_1_RBT____20191015T120946_20191015T121246_20191016T160700_0179_050_223_1980_LN2_O_NT_003.hdf
S3A_SL_1_RBT____20191015T121546_20191015T121846_20191016T160934_0179_050_223_2340_LN2_O_NT_003.hdf
S3A_SL_1_RBT____20191017T130723_20191017T131023_20191018T181427_0179_050_252_2520_LN2_O_NT_003.hdf
S3A_SL_1_RBT____20191017T194820_20191017T195120_20191019T014123_0180_050_256_2340_LN2_O_NT_003.hdf
S3A_SL_1_RBT____20191018T122912_20191018T123212_20191019T172627_0180_050_266_1800_LN2_O_NT_003.hdf
S3A_SL_1_RBT____20191019T120301_20191019T120601_20191020T172539_0179_050_280_1800_LN2_O_NT_003.hdf
S3A_SL_1_RBT____20191019T120601_20191019T120901_20191020T172659_0179_050_280_1980_LN2_O_NT_003.hdf
S3A_SL_1_RBT____20191019T121201_20191019T121501_20191020T172947_0179_050_280_2340_LN2_O_NT_003.hdf
S3A_SL_1_RBT____20191020T113650_20191020T113950_20191021T155904_0179_050_294_1800_LN2_O_NT_003.hdf
S3A_SL_1_RBT____20191020T113950_20191020T114250_20191021T160028_0179_050_294_1980_LN2_O_NT_003.hdf
S3A_SL_1_RBT____20191022T122528_20191022T122828_20191023T162316_0179_050_323_1800_LN2_O_NT_003.hdf
S3A_SL_1_RBT____20191022T122828_20191022T123128_20191023T162424_0179_050_323_1980_LN2_O_NT_003.hdf
S3A_SL_1_RBT____20191023T115917_20191023T120217_20191024T181537_0179_050_337_1800_LN2_O_NT_003.hdf
S3A_SL_1_RBT____20191023T120217_20191023T120517_20191024T181700_0179_050_337_1980_LN2_O_NT_003.hdf
S3A_SL_1_RBT____20191023T120817_20191023T121117_20191024T181932_0179_050_337_2340_LN2_O_NT_003.hdf
S3A_SL_1_RBT____20191024T113606_20191024T113906_20191025T164619_0180_050_351_1980_LN2_O_NT_003.hdf
S3A_SL_1_RBT____20191024T114206_20191024T114506_20191025T164855_0179_050_351_2340_LN2_O_NT_003.hdf
S3A_SL_1_RBT____20191026T122143_20191026T122443_20191027T165912_0179_050_380_1800_LN2_O_NT_003.hdf
S3A_SL_1_RBT____20191026T122443_20191026T122743_20191027T170026_0179_050_380_1980_LN2_O_NT_003.hdf
S3A_SL_1_RBT____20191026T123343_20191026T123643_20191027T170426_0180_050_380_2520_LN2_O_NT_003.hdf
S3A_SL_1_RBT____20191027T115532_20191027T115832_20191028T171044_0179_051_009_1800_LN2_O_NT_003.hdf
S3A_SL_1_RBT____20191027T115832_20191027T120132_20191028T171322_0179_051_009_1980_LN2_O_NT_003.hdf
S3A_SL_1_RBT____20191027T120432_20191027T120732_20191028T171546_0179_051_009_2340_LN2_O_NT_003.hdf
S3A_SL_1_RBT____20191028T131020_20191028T131320_20191029T181138_0179_051_024_1800_LN2_O_NT_003.hdf
S3A_SL_1_RBT____20191029T124409_20191029T124709_20191030T172048_0179_051_038_1800_LN2_O_NT_003.hdf
S3A_SL_1_RBT____20191029T125609_20191029T125909_20191030T172609_0179_051_038_2520_LN2_O_NT_003.hdf
S3A_SL_1_RBT____20191001T230116_20191001T230416_20191003T022222_0179_050_030_0900_LN2_O_NT_003.hdf
S3A_SL_1_RBT____20191002T223505_20191002T223805_20191004T031930_0179_050_044_0900_LN2_O_NT_003.hdf
S3A_SL_1_RBT____20191003T220854_20191003T221154_20191005T030445_0179_050_058_0900_LN2_O_NT_003.hdf
S3A_SL_1_RBT____20191003T234353_20191003T234653_20191005T041723_0179_050_059_0540_LN2_O_NT_003.hdf
S3A_SL_1_RBT____20191004T231742_20191004T232042_20191006T041726_0179_050_073_0540_LN2_O_NT_003.hdf
S3A_SL_1_RBT____20191005T225731_20191005T230031_20191007T024922_0179_050_087_0900_LN2_O_NT_003.hdf
S3A_SL_1_RBT____20191006T223121_20191006T223421_20191008T021301_0179_050_101_0900_LN2_O_NT_003.hdf
S3A_SL_1_RBT____20191007T014419_20191007T014719_20191008T061144_0179_050_103_0360_LN2_O_NT_003.hdf
S3A_SL_1_RBT____20191007T202711_20191007T203011_20191009T155246_0179_050_114_1080_LN2_O_NT_003.hdf
S3A_SL_1_RBT____20191007T220510_20191007T220810_20191010T073003_0179_050_115_0900_LN2_O_NT_003.hdf
S3A_SL_1_RBT____20191008T114503_20191008T114803_20191009T162144_0179_050_123_1620_LN2_O_NT_003.hdf
S3A_SL_1_RBT____20191009T093753_20191009T094053_20191010T151007_0179_050_136_1620_LN2_O_NT_003.hdf
S3A_SL_1_RBT____20191009T111853_20191009T112153_20191010T171717_0179_050_137_1620_LN2_O_NT_003.hdf
S3A_SL_1_RBT____20191010T071243_20191010T071543_20191011T114643_0179_050_149_0540_LN2_O_NT_003.hdf
S3A_SL_1_RBT____20191010T105242_20191010T105542_20191011T160548_0179_050_151_1620_LN2_O_NT_003.hdf
S3A_SL_1_RBT____20191011T014035_20191011T014335_20191012T060733_0179_050_160_0360_LN2_O_NT_003.hdf
S3A_SL_1_RBT____20191011T102631_20191011T102931_20191012T152504_0180_050_165_1620_LN2_O_NT_003.hdf
S3A_SL_1_RBT____20191011T184527_20191011T184827_20191013T002531_0179_050_170_1260_LN2_O_NT_003.hdf
S3A_SL_1_RBT____20191012T181916_20191012T182216_20191013T230919_0179_050_184_1260_LN2_O_NT_003.hdf
S3A_SL_1_RBT____20191013T172106_20191013T172406_20191014T215723_0179_050_197_5400_LN2_O_NT_003.hdf
S3A_SL_1_RBT____20191013T175305_20191013T175605_20191014T222812_0180_050_198_1260_LN2_O_NT_003.hdf
S3A_SL_1_RBT____20191013T225003_20191013T225303_20191015T030239_0179_050_201_0900_LN2_O_NT_003.hdf
S3A_SL_1_RBT____20191014T222352_20191014T222652_20191016T015710_0179_050_215_0900_LN2_O_NT_003.hdf
S3A_SL_1_RBT____20191015T102247_20191015T102547_20191016T142544_0179_050_222_1620_LN2_O_NT_003.hdf
S3A_SL_1_RBT____20191015T120346_20191015T120646_20191016T160531_0179_050_223_1620_LN2_O_NT_003.hdf
S3A_SL_1_RBT____20191015T215741_20191015T220041_20191017T010323_0179_050_229_0900_LN2_O_NT_003.hdf
S3A_SL_1_RBT____20191016T075436_20191016T075736_20191017T123959_0179_050_235_0360_LN2_O_NT_003.hdf
S3A_SL_1_RBT____20191016T095636_20191016T095936_20191017T152606_0179_050_236_1620_LN2_O_NT_003.hdf
S3A_SL_1_RBT____20191016T195331_20191016T195631_20191018T012929_0179_050_242_1080_LN2_O_NT_003.hdf
S3A_SL_1_RBT____20191017T004429_20191017T004729_20191018T060449_0179_050_245_0360_LN2_O_NT_003.hdf
S3A_SL_1_RBT____20191018T222008_20191018T222308_20191020T032233_0179_050_272_0900_LN2_O_NT_003.hdf
S3A_SL_1_RBT____20191019T120001_20191019T120301_20191020T172512_0179_050_280_1620_LN2_O_NT_003.hdf
S3A_SL_1_RBT____20191019T215357_20191019T215657_20191021T014757_0179_050_286_0900_LN2_O_NT_003.hdf
S3A_SL_1_RBT____20191020T194947_20191020T195247_20191022T003158_0179_050_299_1080_LN2_O_NT_003.hdf
S3A_SL_1_RBT____20191021T004044_20191021T004344_20191022T051308_0180_050_302_0360_LN2_O_NT_003.hdf
S3A_SL_1_RBT____20191023T115617_20191023T115917_20191024T181537_0180_050_337_1620_LN2_O_NT_003.hdf
S3A_SL_1_RBT____20191023T201213_20191023T201513_20191025T014837_0179_050_342_1080_LN2_O_NT_003.hdf
S3A_SL_1_RBT____20191023T215012_20191023T215312_20191025T023107_0179_050_343_0900_LN2_O_NT_003.hdf
S3A_SL_1_RBT____20191025T003700_20191025T004000_20191026T054924_0180_050_359_0360_LN2_O_NT_003.hdf
S3A_SL_1_RBT____20191027T101133_20191027T101433_20191028T144445_0179_051_008_1620_LN2_O_NT_003.hdf
S3A_SL_1_RBT____20191027T214628_20191027T214928_20191029T013217_0179_051_015_0900_LN2_O_NT_003.hdf
S3A_SL_1_RBT____20191029T003315_20191029T003615_20191030T050058_0179_051_031_0360_LN2_O_NT_003.hdf
S3A_SL_1_RBT____20191029T223505_20191029T223805_20191031T015742_0179_051_044_0900_LN2_O_NT_003.hdf
S3A_SL_1_RBT____20191030T220854_20191030T221154_20191101T025713_0179_051_058_0900_LN2_O_NT_003.hdf
S3A_SL_1_RBT____20191031T231742_20191031T232042_20191102T034613_0179_051_073_0540_LN2_O_NT_003.hdf
\end{verbatim}

\end{document}